\documentclass[english,12pt]{article}

\usepackage{iftex}
\ifPDFTeX
  \usepackage[T1]{fontenc}
  \usepackage[utf8]{inputenc}
  \usepackage{newtxtext}
\else
  \usepackage[T1]{fontenc}
  \usepackage[utf8]{inputenc}
  \usepackage{fontspec}
\fi
\usepackage{endnotes}
\usepackage{authblk}
\usepackage{geometry}
\usepackage{babel}
\usepackage{amsthm}
\usepackage{amsmath}
\usepackage{amssymb}
\usepackage{mathrsfs}
\usepackage{mathtools}
\usepackage{bbm}
\usepackage{babel}
\usepackage{harvard}
\usepackage{fnpct}
\usepackage{appendix}
\usepackage{pdfpages}
\usepackage[open,openlevel=1]{bookmark}
\usepackage{xcolor}
\usepackage{titling}
\hypersetup{
colorlinks,
linkcolor={red!50!black},
citecolor={blue!50!black},
urlcolor={blue!80!black}
}
\usepackage{multirow}
\usepackage{tikz}
\usepackage{thmtools, thm-restate}  % tools for organizing theorem-like environment
\declaretheoremstyle[
spaceabove=6pt, spacebelow=6pt,
headfont=\normalfont\bfseries,
notefont=\mdseries, notebraces={(}{)},
bodyfont=\normalfont,
postheadspace=1em,
qed=$\blacksquare$
]{examplestyle}

\declaretheoremstyle[
spaceabove=6pt, spacebelow=6pt,
headfont=\normalfont\bfseries,
notefont=\mdseries, notebraces={(}{)},
bodyfont=\itshape,
postheadspace=1em
]{theorem}

\declaretheoremstyle[
spaceabove=6pt, spacebelow=6pt,
headfont=\normalfont\bfseries,
notefont=\mdseries, notebraces={(}{)},
bodyfont=\normalfont,
postheadspace=1em
]{assumption}

\declaretheoremstyle[
spaceabove=4pt, spacebelow=4pt,
headfont=\itshape\bfseries,
notefont=\mdseries, notebraces={(}{)},
bodyfont=\itshape,
postheadspace=0.2em,
qed=\qedsymbol
]{remark}

\declaretheorem[style=theorem]{theorem}
\declaretheorem[style=theorem, numbered=no, name=Theorem]{theorem*}

\declaretheorem[style=assumption,name=Assumption]{assumption}

\declaretheorem[numbered=no,style=definition,name=Question]{question*}  % question
\declaretheorem[style=definition,name=Definition, numbered=no]{definition*}  % definition

\declaretheorem[style=examplestyle]{example}
\renewcommand\thmcontinues[1]{continued}

\newtheorem{thm}{Assumption}

\newcommand{\real}{\mathbb{R}}

\newcommand{\expt}{\text{\normalfont E}}
\newcommand{\var}{\text{\normalfont Var}}
\newcommand{\corr}{\text{\normalfont corr}}

\newcounter{steps}

\newcommand\independent{\protect\mathpalette{\protect\independenT}{\perp}}
\def\independenT#1#2{\mathrel{\rlap{$#1#2$}\mkern2mu{#1#2}}}

\usepackage[nodisplayskipstretch]{setspace}
\usepackage{enumitem}
\setenumerate[1]{label=(\roman*)}
\setenumerate[2]{label=\normalfont{(\alph*)}}

\title{A Generalized Control Function Approach to Production Function Estimation\thanks{We are grateful to Aviv Nevo for useful comments and suggestions.}}
\author{Ulrich Doraszelski\\
University of Pennsylvania\thanks{%
Wharton School, Email: doraszelski@wharton.upenn.edu} \and Lixiong Li\\
Johns Hopkins University\thanks{%
Department of Economics, Email: lixiong.li@jhu.edu}}
\date{\today}

\begin{document}
\maketitle
\begin{abstract}
We develop a generalized control function approach to production function estimation. Our approach accommodates settings in which productivity evolves jointly with other unobservable factors such as latent demand shocks and the invertibility assumption underpinning the traditional proxy variable approach fails. We provide conditions under which the output elasticity of the variable input---and hence the markup---is nonparametrically point-identified. A Neyman orthogonal moment condition ensures oracle efficiency of our GMM estimator. A Monte Carlo exercise shows a large bias for the traditional approach that decreases rapidly and nearly vanishes for our generalized control function approach.
\end{abstract}
\thispagestyle{empty}

\newpage
\setcounter{page}{1}

As the difference of revenue and cost, a firm's profit depends on the demand that it faces and on its production technology. Demand and cost are linked through product innovations, foreign market access, and the firm's advertising decisions. For example, a firm may move downmarket with products that are cheaper to produce but appeal to a broader set of consumers or it may move upmarket with products that are more costly to produce but appeal to more quality-conscious foreign consumers. Cost also depends directly on the firm's R\&D decisions that, in turn, may be subject to financial constraints.

In this paper, we estimate the production function that describes the firm's production technology. To capture the interrelation of demand and cost, we allow the productivity $\omega_{it}$ of firm $i$ in period $t$ to evolve jointly with an unobservable factor $\delta_{it}$. Besides latent demand shocks, $\delta_{it}$ may capture unobserved variation in investment opportunities or financial constraints across firms or time. 

Our setting poses two difficulties for the traditional proxy variable approach to production function estimation developed by \citeasnoun{OLLE:96}, \citeasnoun{LEVI:03}, and \citeasnoun{ACKE:15} (henceforth, OP, LP, and ACF). First, the OP/LP/ACF framework is underpinned by an invertibility assumption that in particular requires that there is no other unobservable, such as our $\delta_{it}$, besides productivity $\omega_{it}$ (the scalar unobservable assumption in ACF). Second, this framework assumes that productivity is governed by a Markov process with law of motion $\omega_{it}=g(\omega_{it-1})+\xi_{it}$. In contrast, we generalize the law of motion for productivity to $\omega_{it}=g(\omega_{it-1},\delta_{it-1})+\xi_{it}$ to allow $\omega_{it}$ and $\delta_{it}$ to evolve jointly.\footnote{Replacing the autonomous Markov process in the OP/LP/ACF framework with a controlled Markov process is straightforward if the control is observed (see \citeasnoun{DELO:13} for learning-by-exporting and \citeasnoun{DORA:07} for R\&D) but not if the control is unobserved.}

To tackle these difficulties, we develop a generalized control function approach to production function estimation. We show that our approach generalizes the control function already present in the OP/LP/ACF framework. We focus on estimating the production function and make no attempt to identify or estimate the law of motion for productivity. We provide conditions under which the output elasticity of the variable input $v_{it}$---and hence the markup following \citeasnoun{DELO:12}---is nonparametrically point-identified and can be estimated by solving a straightforward GMM problem. A Neyman orthogonal moment condition ensures oracle efficiency of our GMM estimator.

\section{Setup}

Our setup follows \citeasnoun{DORA:25} (henceforth, DL), except that we generalize the law of motion for productivity from $\omega_{it}=g(\omega_{it-1})+\xi_{it}$ to $\omega_{it}=g(\omega_{it-1},\delta_{it-1})+\xi_{it}$, where $g(\omega_{it-1},\delta_{it-1})=\expt\left[\omega_{it}|\omega_{it-1},\delta_{it-1}\right]$ and $\xi_{it}$ is the productivity innovation. For concreteness we think of $\delta_{it}$ as latent demand shocks. Firm $i$ in period $t$ uses inputs $k_{it}$ and $v_{it}$ to produce output $q_{it}$ according to the production function
$$
q_{it}=f(k_{it},v_{it})+\omega_{it}+\varepsilon_{it},
$$
where lower case letters denote logs. Capital $k_{it}$ is a predetermined input that is chosen in period $t-1$ whereas $v_{it}$ is freely variable and decided on in period $t$ after the firm observes $\omega_{it}$ and $\delta_{it}$. The disturbance $\varepsilon_{it}$ sits between the firm's output $q_{it}$ as recorded in the data and the output $q^*_{it}=q_{it}-\varepsilon_{it}=f(k_{it},v_{it})+\omega_{it}$ that the firm planned on when it decided on the variable input $v_{it}$. It can be interpreted alternatively as measurement error or as the untransmitted component of productivity.

In what follows, we let $x_{it}=\left(k_{it},v_{it},\ldots\right)$ denote a vector of observables and $z_{it}=\left(k_{it},k_{it-1},v_{it-1},\ldots\right)$ a vector of instruments. As in DL, these vectors can contain additional variables besides the ones listed. 

In the OP/LP/ACF framework, estimation proceeds in two steps. The first step assumes $\expt[\varepsilon_{it}|x_{it}]=0$ and flexibly or nonparametrically estimates the conditional expectation $\expt[q_{it}|x_{it}]$. The second step assumes $\expt[\xi_{it}+\varepsilon_{it}|z_{it}]=0$ and  estimates the production function and the law of motion for productivity using GMM.

\paragraph{Notation.}

For any two vectors $a$ and $b$, we write $a\setminus b$ for the elements of $a$ that are not contained in $b$. To avoid clutter, equalities involving random variables and conditional expectations are understood to hold almost surely. Proofs are deferred to the Supplemental Appendix. 

\section{Results}

We choose the {\em special instrument} $z^s_{it}\subseteq z_{it}$ and partition $z_{it}$ as $z_{it}=(z^s_{it}, z^c_{it})$. In the following assumption, $z^c_{it}$ serve as control variables whereas the special instrument $z^s_{it}$ is excluded from the control function. 

\begin{assumption}\label{assu:exclusion}
$\expt\left[\left.\omega_{it} + \varepsilon_{it} -h\left(z^c_{it}, \expt[q_{it-1}|x_{it-1}]\right) \right| z_{it},\expt[q_{it-1}|x_{it-1}]\right] = 0$ for the control function $h\left(z^c_{it}, \expt[q_{it-1}|x_{it-1}]\right)$.
\end{assumption}

The special instrument $z^s_{it}$ can consist of one or several components of $z_{it}$. As our notation for the control function $h\left(z^c_{it},\expt\left[q_{it-1}|x_{it-1}\right]\right)$ emphasizes, Assumption~\ref{assu:exclusion} requires that conditional on $\left(z^c_{it}, \expt[q_{it-1}|x_{it-1}]\right)$, the special instrument $z^s_{it}$ is uncorrelated with the sum of the transmitted and untransmitted components of productivity $\omega_{it}+\varepsilon_{it}$. In contrast, Assumption~\ref{assu:exclusion} allows the control variables $z^c_{it}$ to be correlated with productivity. 

Because it controls for a part of the variation in productivity, Assumption \ref{assu:exclusion} is less demanding than the condition $\expt[\omega_{it}+\varepsilon_{it}|z_{it}, \expt[q_{it-1}|x_{it-1}]]=0$ used in a conventional instrumental variables approach to production function estimation. It is widely understood that finding instruments in the latter approach is difficult, or perhaps even impossible, in practice \cite{GRIL:98}. 

We further illustrate Assumption \ref{assu:exclusion} with two examples. Example \ref{example:proxy_variable_approach} shows that Assumption~\ref{assu:exclusion} holds in the OP/LP/ACF framework.
\begin{example}[OP/LP/ACF framework]\label{example:proxy_variable_approach}
In the OP/LP/ACF framework, the law of motion for productivity is $\omega_{it}=g(\omega_{it-1})+\xi_{it}$. The GMM estimation in the second step uses the moment condition\footnote{In the OP/LP/ACF framework, $\expt[q_{it-1}|x_{it-1}]$ can be used as an additional instrument. Because the invertibility assumption ensures  $\omega_{it-1}=\expt\left[q_{it-1}|x_{it-1}\right]-f(k_{it-1},v_{it-1})$ and the timing and Markov process assumptions further ensure $\expt[\xi_{it} + \varepsilon_{it}| z_{it}, \omega_{it-1}] = 0$, we have $\expt\left[\xi_{it}+\varepsilon_{it}|z_{it},\expt[q_{it-1}|x_{it-1}]\right]=0$.}
\begin{equation}\label{eq:strong_exclusion}
  \expt\left[\left.q_{it} -f(k_{it},v_{it}) - g\left(\expt\left[q_{it-1}|x_{it-1}\right]-f(k_{it-1},v_{it-1})\right)\right| z_{it},\expt[q_{it-1}|x_{it-1}]\right] = 0.
\end{equation}
Because $q_{it}-f(k_{it},v_{it})=\omega_{it}+\varepsilon_{it}$, moment condition \eqref{eq:strong_exclusion} implies Assumption \ref{assu:exclusion} if we choose $z^c_{it}=\left(k_{it-1},v_{it-1}\right)$, $z^s_{it}=z_{it}\setminus z^c_{it}$, and $h\left(z^c_{it},\expt\left[q_{it-1}|x_{it-1}\right]\right)=g\left(\expt\left[q_{it-1}|x_{it-1}\right]-f(k_{it-1},v_{it-1})\right)$. 

Note that in the OP/LP/ACF framework the invertibility assumption ensures that lagged productivity $\omega_{it-1}=\expt\left[q_{it-1}|x_{it-1}\right]-f(k_{it-1},v_{it-1})$ can be recovered from observables. Moment condition \eqref{eq:strong_exclusion} and Assumption~\ref{assu:exclusion} are therefore closely related. The difference is that moment condition \eqref{eq:strong_exclusion} restricts the control function to take on the specific form implied by the Markov process and invertibility assumptions while Assumption~\ref{assu:exclusion} leaves it unrestricted.
\end{example}

Example \ref{example:proxy_variable_approach} clarifies that the OP/LP/ACF framework can be seen as a control function approach. A key insight of OP is that all that remains of current productivity $\omega_{it}$ after controlling for lagged productivity $\omega_{it-1}$ via the control function $h\left(z^c_{it},\expt\left[q_{it-1}|x_{it-1}\right]\right)=g\left(\expt\left[q_{it-1}|x_{it-1}\right]-f(k_{it-1},v_{it-1})\right)$ is the productivity innovation $\xi_{it}$. Because $\xi_{it}$ is an independent shock, this facilitates finding instruments. 

Following DL, Assumption \ref{assu:exclusion} can also be implied without the invertibility assumption if we choose $x_{it-1}=z_{it}$ in addition to  $z^c_{it}=\left(k_{it-1},v_{it-1}\right)$ and $z^s_{it}=z_{it}\setminus z^c_{it}$. Without invertibility, however, the control function $h\left(z^c_{it},\expt\left[q_{it-1}|x_{it-1}\right]\right)$ generally no longer equals the law of motion $g\left(\expt\left[q_{it-1}|x_{it-1}\right]-f(k_{it-1},v_{it-1})\right)$. DL maintain that the law of motion for productivity is $\omega_{it}=g(\omega_{it-1})+\xi_{it}$.

Relative to the OP/LP/ACF framework and DL, Assumption \ref{assu:exclusion} generalizes the control function approach to production function estimation by allowing us to add variables to the control function. As the conditioning set $\left(z^c_{it}, \expt[q_{it-1}|x_{it-1}]\right)$ contains more variables and thus controls for a larger part of the variation in productivity, the special instrument $z^s_{it}$ contains fewer variables. Adding variables to the control function therefore makes Assumption \ref{assu:exclusion} less demanding. This increases robustness.

Perhaps even more importantly, Assumption \ref{assu:exclusion} allows us to go beyond the OP/LP/ACF framework by not relying on invertibility and accommodating the joint evolution of productivity $\omega_{it}$ and latent demand shocks $\delta_{it}$. It shifts the focus from the law of motion for productivity to specifying a model for the special instrument $z^s_{it}$. As the researcher chooses the special instrument, one can leverage institutional features or auxiliary data to justify Assumption~\ref{assu:exclusion}. Example \ref{example:exogenous_price_shocks} illustrates this point.

\begin{example}[Independent input price shocks]\label{example:exogenous_price_shocks}
Assume that the price of the variable input $p^V_{it}$ evolves according to $p^V_{it} = \kappa(p^V_{it-1}, \eta_{it},\tau_i)$ for some function $\kappa$, where $\eta_{it}$ is a firm- and time-specific input price shock and $\tau_i$ is a firm-specific shifter such as the type of inputs the firm uses. The shifter $\tau_i$ may be correlated with $\delta_{it-1}$ in the law of motion for productivity. This accommodates settings in which the type of inputs links productivity and demand.

Assume that $\eta_{it}$ is an independent shock and therefore in particular independent of $(\xi_{it}, \varepsilon_{it}, \varepsilon_{it-1})$ and of any variables known or chosen by firm in period $t-1$. Because $\omega_{it} = g(\omega_{it-1}, \delta_{it-1}) + \xi_{it}$, it follows that
\begin{equation*}
  (\omega_{it} + \varepsilon_{it}, \omega_{it-1} + \varepsilon_{it-1}) \independent \eta_{it} \big|k_{it}, k_{it-1}, v_{it-1}, \left(p^V_{it'}\right)_{t' < t}.
\end{equation*}
Assume further that the shifter $\tau_i$ can be identified from $\left(p^{V}_{it'}\right)_{t' < t}$.\footnote{Alternatively, assume that we have other information that pins down $\tau_i$. For example, $\tau_i$ may be the location of the firm if there are regional differences in input markets or it may be the countries from which the firm imports inputs.} It follows that
\begin{equation*}
(\omega_{it} + \varepsilon_{it}, \omega_{it-1} + \varepsilon_{it-1}) \independent p^V_{it} \big|k_{it}, k_{it-1}, v_{it-1}, \left(p^V_{it'}\right)_{t' < t}.
\end{equation*}
Choosing $z_{it} = \left(k_{it}, k_{it-1}, v_{it-1}, p^V_{it}, \left(p^{V}_{it'}\right)_{t' < t}\right)$, $x_{it-1} = z_{it}$, $z^s_{it} = p^V_{it}$, and $z^c_{it}=z_{it}\setminus z^s_{it}$, we have
\begin{equation}\label{eq:cond_independent_example_1}
(\omega_{it} + \varepsilon_{it}, \omega_{it-1} + \varepsilon_{it-1}) \independent z^s_{it} \big|z^c_{it}.
\end{equation}
Because $\omega_{it-1} + \varepsilon_{it-1} \independent z^s_{it} \big| z^c_{it}$, we know that $q_{it-1} \independent z^s_{it}\big| z^c_{it}$ and thus that $\expt[q_{it-1}|x_{it-1}]$ depends only on $z^c_{it}$. Equation \eqref{eq:cond_independent_example_1} therefore implies
\begin{equation*}
\omega_{it} + \varepsilon_{it} \independent z^s_{it} \big| z^c_{it}, \expt[q_{it-1}|x_{it-1}]
\end{equation*}
and Assumption \ref{assu:exclusion} holds for the control function $h\left(z^c_{it},\expt[q_{it-1}|x_{it-1}]\right) = \expt\left[\omega_{it} + \varepsilon_{it} | z^c_{it},\expt[q_{it-1}|x_{it-1}]\right]$. Furthermore, because $\expt[q_{it-1}|x_{it-1}]$ depends only on $z^c_{it}$, the control function simplifies from $h(z^c_{it}$, $\expt[q_{it-1}|x_{it-1}])$ to $h(z^c_{it})$.
\end{example}

Assumption \ref{assu:exclusion} ensures that the moment condition
\begin{equation}\label{eq:new_moment}
  \expt\left[\left.q_{it} - f(k_{it}, v_{it}) - h\left(z^c_{it}, \expt[q_{it-1}|x_{it-1}]\right) \right| z_{it}, \expt[q_{it-1}|x_{it-1}]\right] = 0
\end{equation}
holds at the true production function.

We are interested in the output elasticity $\frac{\partial f(k_{it},v_{it})}{\partial v_{it}}$ of the variable input $v_{it}$ as the key to estimating the markup. The choice of the special instrument $z^s_{it}$ entails a tradeoff. As noted above, as the conditioning set $\left(z^c_{it}, \expt[q_{it-1}|x_{it-1}]\right)$ contains more variables and thus controls for a larger part of the variation in productivity, the special instrument $z^s_{it}$ contains fewer variables. This leaves less exogenous variation in $z^s_{it}$ for identifying $\frac{\partial f(k_{it},v_{it})}{\partial v_{it}}$. To ensure that $z^s_{it}$ has sufficient identification power, we make the following assumption:

\begin{assumption}\label{assu:complete_cond1}
Let $k_{it}\in  z^c_{it}$. Conditional on $\left(z^c_{it},\expt[q_{it-1}|x_{it-1}]\right)$, $z^s_{it}$ is a complete instrument for $v_{it}$.
\end{assumption}

Assumption~\ref{assu:complete_cond1} places restrictions on the underlying economic model. In particular, it rules out the case where the law of motion for productivity is $\omega_{it}=g(\omega_{it-1})+\xi_{it}$ and the variable input demand $v_{it}=\kappa(k_{it},\omega_{it})$ is some function $\kappa$ that depends only on capital and productivity and can be inverted for productivity.\footnote{Assumption~\ref{assu:complete_cond1} therefore also rules out the nonidentification result in \citeasnoun{GAND:20}.} In this case, for any choice of the special instrument $z^s_{it}\subseteq z_{it} \setminus (k_{it}, k_{it-1}, v_{it-1})$,  $z^s_{it}$ is independent of $\omega_{it}$ conditional on $\left(z^c_{it}, \expt[q_{it-1}|x_{it-1}]\right)$ because $\omega_{it-1}$ is pinned down by $\left(z^c_{it}, \expt[q_{it-1}|x_{it-1}]\right)$ and the productivity innovation $\xi_{it}$ is an independent shock. Consequently, $z^s_{it}$ cannot generate any variation in $v_{it} = \kappa(k_{it}, \omega_{it})$ after controlling for $\left(z^c_{it}, \expt[q_{it-1}|x_{it-1}]\right)$. Note that this lack-of-variation argument breaks down if latent demand shocks $\delta_{it}$ enter the law of motion for productivity or the variable input demand. Hence, latent demand shocks $\delta_{it}$ make room for Assumption~\ref{assu:complete_cond1} to hold.

Our main identification result is the following:
\begin{theorem}\label{thm:identification}
Under Assumptions \ref{assu:exclusion} and \ref{assu:complete_cond1}, moment condition \eqref{eq:new_moment} nonparametrically point-identifies $\frac{\partial f(k_{it},v_{it})}{\partial v_{it}}$.
\end{theorem}

Turning from identification to estimation, for any weighting function $\varphi(z_{it})$ of the instruments $z_{it}$, moment condition \eqref{eq:new_moment} implies that the moment condition
\begin{equation}\label{eq:uncond_new_moment}
  \expt\left[ \varphi(z_{it})\left(q_{it} - f(k_{it}, v_{it}) - h\left(z^c_{it}, \expt[q_{it-1}|x_{it-1}]\right)\right)\right] = 0
\end{equation}
holds at the true production function. GMM estimation based on moment condition \eqref{eq:uncond_new_moment} can be conducted by viewing the control function $h\left(z^c_{it}, \expt[q_{it-1}|x_{it-1}]\right)$ as a nuisance parameter and estimating it alongside the production function, as in the OP/LP/ACF framework.\footnote{The law of motion for productivity cannot generally be recovered from estimates of the production and control functions.} However, depending on the number of control variables $z^c_{it}$, the dimension of the control function may be high. This increases the asymptotic variance of the estimates and can create numerical challenges for minimizing the GMM objective function.

Our main estimation result shows that these drawbacks can be avoided:
\begin{theorem}\label{thm:orth_uncond_new_moment}
For any weighting function $\varphi(z_{it})$ of the instruments $z_{it}$, moment condition \eqref{eq:new_moment} implies that the moment condition
\begin{equation}\label{eq:orth_uncond_new_moment}
  \expt\left[\left(\varphi(z_{it}\right) - \tilde{\varphi}_{it}) \left(q_{it} - f(k_{it}, v_{it}) - \left(\tilde{q}_{it} - \tilde{f}_{it}\right)\right)\right] = 0,
\end{equation}
where
\begin{gather*}
\tilde{\varphi}_{it} =\expt\left[\varphi(z_{it}) | z^c_{it}, \expt[q_{it-1}|x_{it-1}] \right], \quad
\tilde{q}_{it}-\tilde{f}_{it} =\expt\left[q_{it}-f(k_{it}, v_{it}) | z^c_{it}, \expt[q_{it-1}|x_{it-1}] \right],
\end{gather*}
holds at the true production function. Moreover, if a production function satisfies moment condition \eqref{eq:orth_uncond_new_moment}, then it satisfies moment condition \eqref{eq:uncond_new_moment} for the control function $h\left(z^c_{it}, \expt[q_{it-1}|x_{it-1}]\right) = \tilde{q}_{it} - \tilde{f}_{it}$. Finally, if $\varphi(z_{it})$ and $q_{it}-f(k_{it}, v_{it})$ have finite $L^2$ norms, then
moment condition \eqref{eq:orth_uncond_new_moment} is Neyman orthogonal with respect to $L^2$-integrable perturbations of $\left(\tilde{\varphi}_{it},\tilde{q}_{it}-\tilde{f}_{it}\right)$.
\end{theorem}

Applying results from the double-debiased machine learning literature \cite{CHER:18}, Neyman orthogonality ensures that the GMM estimator based on moment condition \eqref{eq:orth_uncond_new_moment} is oracle efficient as long as the estimators for $\tilde{\varphi}_{it}$, $\tilde{q}_{it}$, and $\tilde{f}_{it}$ converge sufficiently fast. This means that the asymptotic distribution of the GMM estimator is {\em as if} the true values of
$\tilde{\varphi}_{it}$ and $\tilde{q}_{it}-\tilde{f}_{it}$ are known.

A complication arises because estimating $\tilde{\varphi}_{it}$ and $\tilde{q}_{it}-\tilde{f}_{it}$ requires itself a plugin estimator for $\expt[q_{it-1}|x_{it-1}]$. To the best of our knowledge, the literature has not yet developed a treatment for such ``double plugin'' estimators. In what follows, we therefore focus on the case where the control function simplifies from $h\left(z^c_{it},\expt[q_{it-1}|x_{it-1}]\right)$ to $h\left(z^c_{it}\right)$ as in Example \ref{example:exogenous_price_shocks}. Thus, the plugin estimator for $\expt[q_{it-1}|x_{it-1}]$ is no longer required, and $\tilde{\varphi}_{it} =\expt\left[\left.\varphi(z_{it}) \right| z^c_{it}\right]$ and $\tilde{q}_{it}-\tilde{f}_{it} =\expt\left[\left.q_{it}-f(k_{it}, v_{it}) \right| z^c_{it}\right]$.

While our generalized control function approach accommodates the joint evolution of productivity $\omega_{it}$ and latent demand shocks $\delta_{it}$, it is not costless. If moment condition \eqref{eq:strong_exclusion} holds, then for any choice of the special instrument $z^s_{it}\subsetneq z_{it}\setminus (k_{it-1}, v_{it-1})$ it is less efficient than the OP/LP/ACF procedure. This reflects a general tradeoff between robustness and efficiency in nonparametric estimation. Our generalized control function approach makes this tradeoff explicit by allowing a researcher to target greater robustness (smaller $z^s_{it}$) or greater efficiency (larger $z^s_{it}$).

\section{Monte Carlo Exercise}

\paragraph{Data generating process.}

Similar to DL, we specify the CES production function
\begin{equation*}
f(k_{it},v_{it})=\frac{\nu}{\rho}\ln\left(\alpha\exp(\rho k_{it})+(1-\alpha)\exp(\rho v_{it})\right)
\end{equation*}
with $\alpha=0.3$, $\rho=-1$, and $\nu=0.95$, the disturbance $\varepsilon_{it}\sim N\left(0,0.5^2\right)$, and the CES demand $q^*_{it}=\delta_{1it}-(1+\exp(-\delta_{2it}))p_{it}$, where $p_{it}$ is the output price and $\delta_{it}=(\delta_{1it},\delta_{2it})$ captures shocks to the demand the firm faces and unobserved rivals.

Different from DL, the price of capital $p^K_{it}$, the price of the variable input $p^V_{it}$, and the latent demand shocks $\delta_{1it}$ and $\delta_{2it}$ follow Gaussian $AR(1)$ processes. We parameterize these processes so that $\expt[p^K_{it}]=\expt[p^V_{it}]=0$, $\expt[\delta_{1it}]=10$, $\expt[\delta_{2it}]=-1.3543$, $\var(p^K_{it})=\var(p^V_{it})=\var(\delta_{2it})=0.5^2$, $\var(\delta_{1it})=5^2$, and the autocorrelation is $0.7$. Short-run profit maximization implies the markup $\mu_{it}=\frac{P_{it}}{MC_{it}}=1+\exp(\delta_{2i})$, where $MC_{it}$ is marginal cost, and thus $\expt\left[\ln\mu_{it}\right]=0.25$ and $\var\left(\ln\mu_{it}\right)=0.0126$.

We specify the law of motion for productivity
\begin{equation*}
g(\omega_{it-1},\delta_{it-1})=\mu_\omega+\rho_\omega\omega_{it-1}+\rho_{\delta_1}\delta_{1it-1}+\rho_{\delta_2}\delta_{2it-1}
\end{equation*}
and the productivity innovation $\xi_{it}\sim N\left(0,\sigma^2_{\omega}\right)$.
We parameterize $\mu_\omega$, $\rho_\omega$, $\rho_{\delta_1}$, $\rho_{\delta_2}$, and $\sigma^2_{\omega}$ so that $\expt[\omega_{it}]=0$, $\var(\omega_{it})=0.5^2$, $\corr(\omega_{it},\omega_{it-1})=0.7$, $\corr(\omega_{it},\delta_{1it})=0.3$, and
$\corr(\omega_{it},\delta_{2it})=-0.3$. This aligns with the notion that more productive firms participate in larger and more competitive markets.

We simulate $S=1,000$ datasets with $N=5,000$ firms and $T=20$ periods. We refer the reader to DL for further details on the data generating process.

\paragraph{Estimation.}

We use GMM estimation based on moment condition \eqref{eq:orth_uncond_new_moment} to estimate the production function parameters $\theta=\left(\alpha,\rho,\nu\right)$. Our instruments are $z_{it}=\left(k_{it},k_{it-1},v_{it-1},p_{it-1},p^V_{it},p^V_{it-1}\right)$ and our special instrument is $z^s_{it}=p^V_{it}$ as in Example \ref{example:exogenous_price_shocks}. Our weighting function $\varphi(z_{it})$ is the complete set of Hermite polynomials of total degree 4 in the variables in $z_{it}$. We estimate $\tilde{\varphi}_{it}=\expt\left[\varphi(z_{it})|z^c_{it}\right]$ by OLS using the complete set of Hermite polynomials of total degree $d$ in the variables in $z^c_{it}$. We proceed similarly to estimate $\tilde{q}_{it}-\tilde{f}_{it}=\expt\left[q_{it}-f(k_{it},v_{it})|z^c_{it}\right]$. The latter must be re-estimated at each iteration of the GMM problem. Even though the control function $h\left(z^c_{it}\right) = \expt\left[\omega_{it} + \varepsilon_{it} | z^c_{it} \right]$ is absent from moment condition \eqref{eq:orth_uncond_new_moment}, the total degree $d$ implicitly determines how well we can approximate it. We accordingly explore $d\in\{2,3,4,5\}$. We provide further details on the GMM estimator in the Supplemental Appendix.

With an estimate of $\theta$ in hand, we estimate the markup $\mu_{it}$ of firm $i$ in period $t$ as
\begin{equation*}
\ln\mu_{it}+\varepsilon_{it}=p_{it}+q_{it}-p^V_{it}-v_{it}+\ln\frac{\partial f(k_{it},v_{it})}{\partial v_{it}}.
\end{equation*}
The right-hand side is the log of the output elasticity minus the log of the expenditure share of the variable input. Noting that the disturbance $\varepsilon_{it}$ averages out as $\expt[\varepsilon_{it}]=0$, we refer to the average of $\ln\mu_{it}+\varepsilon_{it}$ across firms and time simply as the average log markup.

\paragraph{Results.}

As a baseline, we implement the modified OP/LP/ACF procedure described in DL by choosing $x_{it-1}=z_{it}$ and explicitly including the first-order bias correction. We use a univariate Hermite polynomial of order 4 to approximate the law of motion for productivity. The bias in the average log markup is $0.1277$ (compared to its true value of $0.25$) and the mean squared error is $0.0164$. The large bias is not surprising given that the joint evolution of productivity $\omega_{it}$ and latent demand shocks $\delta_{it}$ is outside the scope of the OP/LP/ACF procedure.

\begin{figure}
\centerline{\includegraphics[scale=0.475]{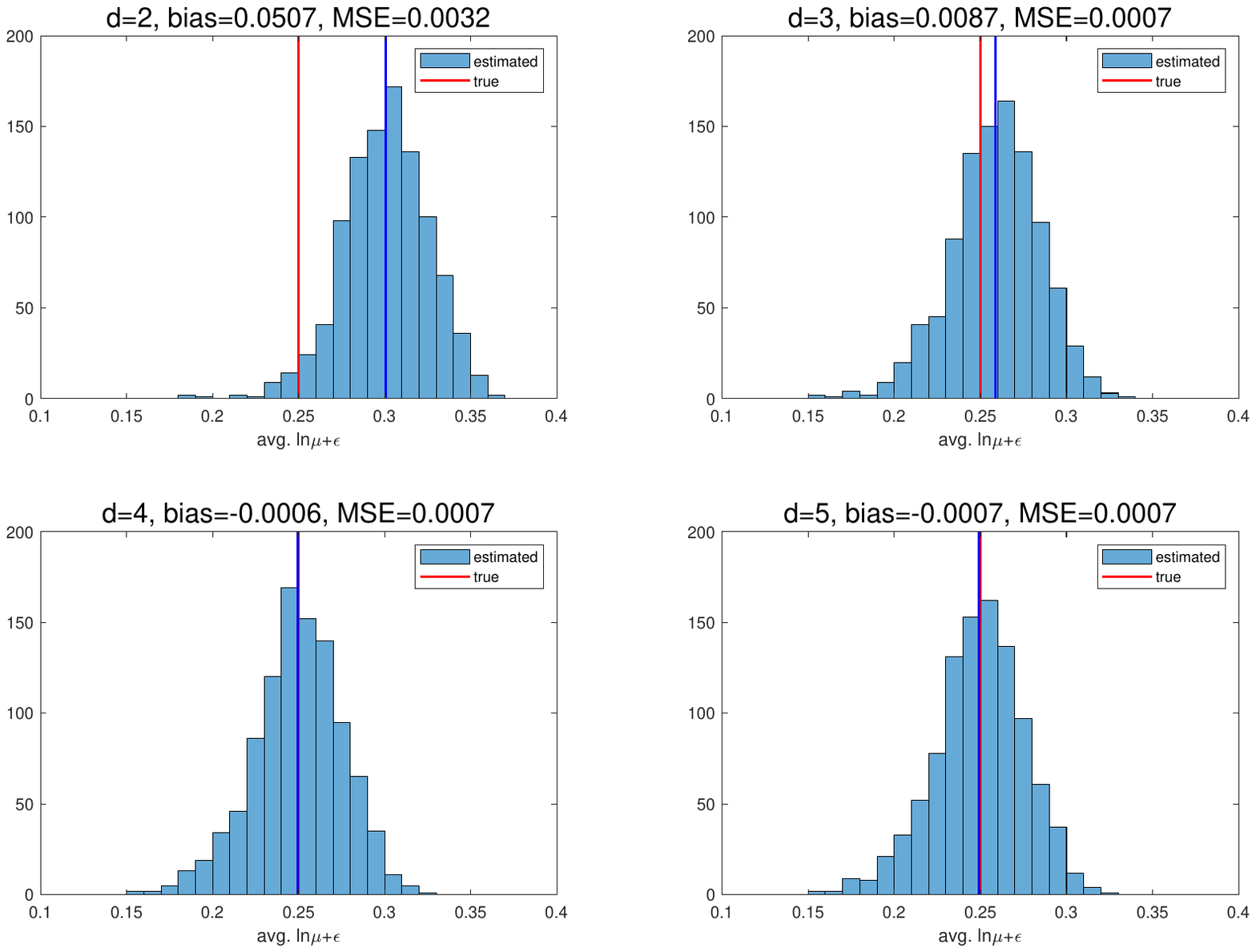}}
\caption{Distribution of average log markup for Hermite polynomials of total degree $d\in\{2,3,4,5\}$.\label{fig:results}}
\end{figure}

Turning to the generalized control function approach in this paper, Figure \ref{fig:results} shows the distribution of the average log markup. As can be seen, the results rapidly improve with the total degree $d$ of the complete set of Hermite polynomials in the variables in $z^c_{it}$. The bias decreases from $0.0507$ for $d=2$ to $-0.0006$ for $d=4$ and the mean squared error from to $0.0032$ to $0.0007$. There are no further improvements going from $d=4$ to $d=5$.

\section{Concluding Remarks}

We provide conditions for consistently estimating the production function in empirically relevant settings that are outside the scope of the OP/LP/ACF framework. As such, our approach complements the OP/LP/ACF framework. It generalizes the control function that is already present in the OP/LP/ACF framework and requires solving a straightforward GMM problem. We hope that it proves valuable for applied researchers seeking to estimate the production function and the markup from it.

\bibliography{references}

\newpage
\title{Supplemental Appendix: A Generalized Control Function Approach to Production Function Estimation}
\author{Ulrich Doraszelski\\
University of Pennsylvania \and Lixiong Li\\
Johns Hopkins University}
\date{\today}
\maketitle
\pagenumbering{roman}

\paragraph{Proof of Theorem \ref{thm:identification}.}

Let $f^0(k_{it},v_{it})$ denote the true production function.  Because $q_{it} - f^0(k_{it}, v_{it}) = \omega_{it} + \varepsilon_{it}$, Assumption \ref{assu:exclusion} implies that $f^0(k_{it},v_{it})$ satisfies moment condition \eqref{eq:new_moment} for the control function $h^0\left(z^c_{it}, \expt[q_{it-1}|x_{it-1}]\right) = \expt\left[\left.q_{it} - f^0(k_{it}, v_{it})\right|z^c_{it}, \expt[q_{it-1}|x_{it-1}]\right]$. Let $\tilde{f}(k_{it},v_{it})$ be some production function that also satisfies moment condition \eqref{eq:new_moment} for some control function $\tilde{h}\left(z^c_{it}, \expt[q_{it-1}|x_{it-1}]\right)$. To show that $\frac{\partial f(k_{it}, v_{it})}{\partial v_{it}}$ is nonparametrically point-identified,  we show that, for any $k_{it}$, the difference $f^0(k_{it}, v_{it}) - \tilde{f}(k_{it}, v_{it})$ does not change with $v_{it}$ almost surely.

Because both $(f^0, h^0)$ and $(\tilde{f}, \tilde{h})$ satisfy moment condition \eqref{eq:new_moment}, taking the difference yields
\begin{eqnarray*}
&&\expt\left[\left.f^0(k_{it}, v_{it}) - \tilde{f}(k_{it}, v_{it}) \right| z_{it}, \expt[q_{it-1}|x_{it-1}]\right] \\
&=&\expt\left[\left.\tilde{h}\left(z^c_{it}, \expt[q_{it-1}|x_{it-1}]\right) - h^0\left(z^c_{it}, \expt[q_{it-1}|x_{it-1}]\right) \right|z_{it}, \expt[q_{it-1}|x_{it-1}]   \right]\\
&=&\tilde{h}\left(z^c_{it}, \expt[q_{it-1}|x_{it-1}]\right) - h^0\left(z^c_{it}, \expt[q_{it-1}|x_{it-1}]\right).
\end{eqnarray*}
This implies the exclusion restriction
\begin{eqnarray*}
&&\expt\left[\left.f^0(k_{it}, v_{it}) - \tilde{f}(k_{it}, v_{it}) \right| z_{it}, \expt[q_{it-1}|x_{it-1}]\right] \\
&=& \expt\left[\left.f^0(k_{it}, v_{it}) - \tilde{f}(k_{it}, v_{it}) \right| z^c_{it}, \expt[q_{it-1}|x_{it-1}]\right].
\end{eqnarray*}

Assumption \ref{assu:complete_cond1} ensures that, for any function $\kappa$, the exclusion restriction
\begin{eqnarray*}
&&\expt\left[\left.\kappa\left(v_{it}, z^c_{it}, \expt[q_{it-1}|x_{it-1}]\right) \right| z_{it}, \expt[q_{it-1}|x_{it-1}] \right] \\
&=& \expt\left[\left.\kappa\left(v_{it}, z^c_{it}, \expt[q_{it-1}|x_{it-1}]\right) \right| z^c_{it}, \expt[q_{it-1}|x_{it-1}] \right]
\end{eqnarray*}
holds only if $\kappa$ does not change with $v_{it}$ almost surely. Recalling that $k_{it}\in z^c_{it}$ and setting $\kappa\left(v_{it}, z^c_{it}, \expt[q_{it-1}|x_{it-1}]\right) =f^0(k_{it}, v_{it}) - \tilde{f}(k_{it}, v_{it})$ therefore establishes the result.

\paragraph{Proof of Theorem \ref{thm:orth_uncond_new_moment}.}

We divide the proof into three parts.

We first prove that moment condition \eqref{eq:new_moment} implies moment condition \eqref{eq:orth_uncond_new_moment}. Recalling the definitions of $\tilde{\varphi}_{it}$ and $\tilde{q}_{it}-\tilde{f}_{it}$ in Theorem \ref{thm:orth_uncond_new_moment}, moment condition \eqref{eq:new_moment} implies
\begin{equation}\label{eq:interim}
\expt\left[\left.q_{it} - f(k_{it}, v_{it}) - \left(\tilde{q}_{it} - \tilde{f}_{it}\right) \right| z_{it}, \expt[q_{it-1}|x_{it-1}]\right] = 0.
\end{equation}
This is because moment condition \eqref{eq:new_moment} implies $h(z^c_{it}, \expt[q_{it-1}|x_{it-1}]) = \expt[q_{it} - f(k_{it}, v_{it})| z_{it}, \expt[q_{it-1}|x_{it-1}]]$ so that the control function has to satisfy $h(z^c_{it}, \expt[q_{it-1}|x_{it-1}]) = \expt[q_{it} - f(k_{it}, v_{it})| z^c_{it}, \expt[q_{it-1}|x_{it-1}]] = \tilde{q}_{it} - \tilde{f}_{it}$.
Because $\varphi(z_{it}) - \tilde{\varphi}_{it}$ is pinned down by $\left(z_{it}, \expt[q_{it-1}|x_{it-1}]\right)$, we have
\begin{eqnarray*}
&& \expt\left[ \left(\varphi(z_{it}) - \tilde{\varphi}_{it}\right) \left(q_{it} - f(k_{it}, v_{it}) - \left(\tilde{q}_{it} - \tilde{f}_{it}\right)\right)\right]\\
&=& \expt\left[ \left(\varphi(z_{it}) - \tilde{\varphi}_{it}\right) \expt\left[\left.q_{it} - f(k_{it}, v_{it}) - \left(\tilde{q}_{it} - \tilde{f}_{it}\right) \right| z_{it}, \expt[q_{it-1}|x_{it-1}]\right]\right]\\
&=& 0,
\end{eqnarray*}
where the first equality is due to the law of iterated expectations and the second equality to equation \eqref{eq:interim}.

Next, we prove that moment condition \eqref{eq:orth_uncond_new_moment} implies moment condition \eqref{eq:uncond_new_moment} with
$h\left(z^c_{it}, \expt[q_{it-1}|x_{it-1}]\right) = \tilde{q}_{it} - \tilde{f}_{it}$.
Note that
\begin{eqnarray}
  &&\expt\left[\tilde{\varphi}_{it} \left(q_{it} - f(k_{it}, v_{it}) - \left(\tilde{q}_{it} - \tilde{f}_{it}\right)\right)\right] \label{eq:temp_middle}\\
  &=&\expt\left[\tilde{\varphi}_{it}\expt\left[\left.q_{it} - f(k_{it}, v_{it}) - \left(\tilde{q}_{it} - \tilde{f}_{it}\right) \right|z^c_{it}, \expt[q_{it-1}|x_{it-1}] \right] \right] \nonumber\\
  &=& 0, \nonumber
\end{eqnarray}
where the first equality is due to the law of iterated expectations and the second equality to $\tilde{q}_{it} - \tilde{f}_{it} =  \expt[q_{it} - f(k_{it}, v_{it}) | z^c_{it}, \expt[q_{it-1}|x_{it-1}]] = 0$ by the definition of $\tilde{q}_{it}-\tilde{f}_{it}$. Adding moment condition \eqref{eq:orth_uncond_new_moment} and moment condition \eqref{eq:temp_middle} implies moment condition \eqref{eq:uncond_new_moment} with $h\left(z^c_{it}, \expt[q_{it-1}|x_{it-1}]\right) = \tilde{q}_{it} - \tilde{f}_{it}$.

Finally, we prove that moment condition \eqref{eq:orth_uncond_new_moment} is Neyman orthogonal. Define the shorthand $\tilde{z}^c_{it}=\left(z^c_{it}, \expt[q_{it-1}|x_{it-1}]\right)$. Define $\mathcal{F}$ to be the set of function tuples $(\eta, \zeta)$ such that $\expt[\eta^2(\tilde{z}^c_{it})] < \infty$ and $\expt[\zeta^2(\tilde{z}^c_{it})] < \infty$. Because $\varphi(z_{it})$ and $q_{it}-f(k_{it}, v_{it})$ have finite $L^2$ norms, for any $(\eta, \zeta)\in \mathcal{F}$ and for any $\lambda = (\lambda_1, \lambda_2)\in \real^2$, we have
\begin{equation*}
  \expt\left| \left(\varphi(z_{it}) - \tilde{\varphi}_{it} - \lambda_1 \eta(\tilde{z}^c_{it})\right)\left(q_{it} - f(k_{it}, v_{it}) - \left(\tilde{q}_{it}  - \tilde{f}_{it} + \lambda_2\zeta(\tilde{z}^c_{it}) \right)\right) \right| < \infty.
\end{equation*}
Moment condition \eqref{eq:orth_uncond_new_moment} is therefore well-defined for any $L^2$-integrable perturbation of $\left(\tilde{\varphi}_{it},\tilde{q}_{it} - \tilde{f}_{it}\right)$. Next, note that
\begin{eqnarray*}
&&\left. \frac{\partial \expt (\varphi(z_{it}) - \tilde{\varphi}_{it} - \lambda_1 \eta(\tilde{z}^c_{it}))(q_{it} - f(k_{it}, v_{it}) - (\tilde{q}_{it} - \tilde{f}_{it} + \lambda_2\zeta(\tilde{z}^c_{it})  ))  }{\partial \lambda}\right|_{\lambda=0}\\
&=&
\begin{pmatrix}
  \expt\left[\eta(\tilde{z}^c_{it})\left(q_{it} - f(k_{it}, v_{it}) - \left(\tilde{q}_{it}  - \tilde{f}_{it}\right)\right)\right]\\
  -\expt\left[(\varphi(z_{it}) - \tilde{\varphi}_{it})\zeta(\tilde{z}^c_{it})\right]
\end{pmatrix}
=
\begin{pmatrix}
0\\
0
\end{pmatrix},
\end{eqnarray*}
where the second equality is due to the law of iterated expectations and the definitions of $\tilde{\varphi}_{it}$ and $\tilde{q}_{it}-\tilde{f}_{it}$.

\paragraph{GMM estimator.}

Corresponding to moment condition \eqref{eq:orth_uncond_new_moment}, define the moment function
\begin{equation}\label{eq:moment_791}
m_{it}(\theta)=q_{it}-f(k_{it},v_{it};\theta)-\left(\tilde{q}_{it}-\tilde{f}_{it}(\theta)\right),
\end{equation}
where we make the parameterization of the production function explicit.

We solve the GMM problem
\begin{equation*}\label{eq:GMM_795}
\min_\theta \left(\frac{1}{NT}\sum_{i,t}\left(\varphi(z_{it})-\tilde{\varphi}_{it}\right)m_{it}(\theta)\right)^\top W \left(\frac{1}{NT}\sum_{i,t}\left(\varphi(z_{it})-\tilde{\varphi}_{it}\right)m_{it}(\theta)\right),
\end{equation*}
where the superscript $\top$ denotes the transpose. We use the weighting matrix
\begin{equation*}
	W=\left(\frac{1}{NT-1}\sum_{i,t}\left(\left(\varphi(z_{it})-\tilde{\varphi}_{it}\right)m_{it}\left(\theta^0\right) - \hat{\mu}\right)^\top\left(\left(\varphi(z_{it})-\tilde{\varphi}_{it}\right)m_{it}\left(\theta^0\right) - \hat{\mu}\right)\right)^{-1},
\end{equation*}
where
\begin{equation*}
\hat{\mu} = \frac{1}{NT}\sum_{i,t}\left(\varphi(z_{it})-\tilde{\varphi}_{it}\right)m_{it}\left(\theta^0\right)
\end{equation*}
and $\theta^0$ denotes the true parameters.

Recall that $\tilde{\varphi}_{it}=\expt\left[\varphi(z_{it})|z^c_{it}\right]$. To the extent that terms in $\varphi(z_{it})$ can be perfectly predicted by the complete set of Hermite polynomials of total degree $d$ in the variables in $z^c_{it}$, the matrix $\Phi$ with rows $\varphi(z_{it})-\tilde{\varphi}_{it}$ is rank deficient. In solving the GMM problem, we therefore use a selection of columns from the matrix $\Phi$ that has full rank. To construct this selection, we start with an empty matrix and keep adding columns from the matrix $\Phi$ as long as this increases the rank.

\end{document}